# Non-Ohmic behavior of carrier transport in highly disordered graphene

Shun-Tsung Lo[1,][*], Chiashain Chuang[2, 3,][*], R. K. Puddy[2], T.-M.Chen[4], C. G. Smith[2], and C.-T. Liang[1, 3]

[1]*Graduate Institute of Applied Physics, National Taiwan University, Taipei 106, Taiwan*
[2]*Cavendish Laboratory, J. J. Thomson Avenue, Cambridge CB3 0HE, United Kingdom*
[3]*Department of Physics, National Taiwan University, Taipei 106, Taiwan*
[4]*Department of Physics, National Cheng-Kung University, Tainan 701, Taiwan*

* These authors contributed equally to this work
Correspondences should be addressed to C.-T. Liang (ctliang@phys.ntu.edu.tw)

## Abstract

We report measurements of disordered graphene probed by both a high electric field and a high magnetic field. By apply a high source-drain voltage $V_{sd}$, we are able to study the current-voltage relation $I$-$V_{sd}$ of our device. With increasing $V_{sd}$, a crossover from the linear $I$-$V_{sd}$ regime to the non-linear one, and eventually to activationless-hopping transport occurs. In the activationless-hopping regime, the importance of Coulomb interactions between charged carriers is demonstrated. Moreover, we show that delocalization of carriers which are strongly localized at low $T$ and at small $V_{sd}$ occurs with the presence of high electric field and perpendicular magnetic field..

I. Introduction

Graphene, a mono-layer of carbon atoms bonded in a hexagonal lattice, is an ideal two-dimensional (2D) system[1-3]. Because of its excellent electrical[1], mechanical[4] and thermal properties[5], graphene continues to attract much interest. The observations of fascinating effects such as the fractional quantum Hall effect[6,7] and ballistic conductance quantization[8,9] require graphene samples of ultra-high quality. Therefore a lot of efforts have been made so as to prepare high-quality graphene. On the other hand, there is a dearth of study of disordered graphene devices which are equally of fundamental importance as well as of potential applications. Moreover, it is important to understand the nature of disordered graphene so as to fully realize its potentials in device applications as well as to provide fundamental understanding of strong localization, hopping, de-phasing in the strongly localized regime, and transport under a high electric field which are not normally observable in pristine graphene samples. More importantly, it has been shown that disordered graphene such as chemically doped graphene[10,11] and graphene nanoribbon[12,13] can have a band gap which is of potential applications in nanoelectronics. Furthermore, graphitic oxide, a formed of disordered graphene-based material, can be used as a gate dielectric[14] for experimental realization of graphene-based transistors. Therefore it is highly desirable to obtain a thorough understanding of the nature of disordered graphene. It is believed that with a sufficient amount of short-range disorder, the transport properties of graphene resemble those of conventional 2D systems due to strong intervalley mixing[15-17].

In a highly disordered system, strong localization of carriers occurs and variable-range hopping (VRH) rather than diffusive motion would then dominate the transport[18-20]. In the limit of very low electric field $E$, the required energy for a hop between localized states is from phonon absorption. Therefore the conductance $G \equiv I/V_{sd}$ is $T$-dependent and insensitive to the variation of the electric field, showing Ohmic behavior. In this regime, the VRH conductance increases exponentially with $T$ as[21-23]

$$G(T) = G_0 \exp[-(T_0/T)^p], \quad (1)$$

with pre-factor $G_0$ and characteristic temperature $T_0$. The exponent $p$ can be used to distinguish whether Coulomb interactions are important. In a 2D system, $p = 1/3$ and $kT_0 = 13.8/g_F\xi^2$ with a constant density of states (DOS) $g_F$ at the Fermi energy $E_F$ for Mott VRH while $p = 1/2$ and $kT_0 = 6.2e^2/4\pi\varepsilon_r\varepsilon_0\xi$ for Efro-Shklovskii (ES) VRH when Coulomb interactions lead to the vanishing of DOS at $E_F$. Here k, e, $\varepsilon_r$, $\varepsilon_0$, and $\xi$ denote Boltzmann constant, electronic charge, dielectric constant, vacuum permittivity, and localization length. With increasing electric field $E$, the energy scale would be reduced by an amount of $eEr_{hop}$ in the field direction and therefore the energy involved in the hopping process comes both from phonon absorption leading to the $T$ dependence of $G$ and electrical energy leading to the $E$ dependence of $G$, where $r_{hop}$ represents the hopping length. In this regime, hopping conductance has the form[24-28]

$$G(E, T) = G(0, T)\exp(e(E-E_{c1})L/kT), \qquad (2)$$

where $G(0, T)$ is a pre-factor and $L$ is a length parameter related to the hopping length according to $L \propto r_{hop}$ [24, 25] or $L \propto r_{hop}^2$ [31]. At high fields $E > E_{c2} \equiv kT/e\xi$ [29, 32], the hopping conduction becomes activationless and the resulting current is expressed as[27, 28, 29, 32, 33]

$$I = I_0 \exp(-(E_0/E)^s), \qquad (3)$$

which is $T$-independent. Here $I_0$ is a pre-factor and $E_0 \sim kT_0/e\xi$. "Activationless" means that the carries participating in the hopping transport acquire energy utterly from the electric field. The exponent $s$ is generally believed to be equal to $p$ but there exists some experimental evidence suggesting that $s$ and $p$ can be different[33]. In this report, both an electric and a magnetic field are utilized to probe the transport features of highly disordered graphene. With increasing electric field, a crossover from activation to activationless transport occurs. Moreover the ES VRH is shown to dominate the transport either in the phonon-assisted or electric-field-assisted hopping regime, indicative of the importance of Coulomb interactions between charged carriers in disordered graphene. Most interestingly, we find that both electric and magnetic fields can introduce delocalization effects in our system.

## II. Experimental details

A single-layer graphene flake, mechanically exfoliated from natural graphite onto a highly doped Si substrate capped with a 300-nm-thick $SiO_2$ layer, were used in our study. The detailed fabrication process can be found in Ref. [34]. In our device, disorder is introduced by exposing the graphene flake to hydrogen plasma. The experiments were performed in a $He^4$ cryostat equipped with a superconducting magnet. Two-terminal current-voltage $I$-$V$ dc measurements at various temperatures $T$ and various magnetic fields $B$ were performed.

## III. Results and discussion

Figure 1 (a) shows that the conductance $G = dI/dV_{sd}$ as a function of temperature $T$ in the zero bias limit (-3 mV $\leq V_{sd} \leq$ 3 mV) and at zero magnetic field can be well described by Eq. (1) with $p = 1/2$ when $T \geq 4.5$ K, revealing that Coulomb interactions play a key role in our system. Moreover we can extract $T_0 \sim$ 69 K from the linear fit in Fig. 1 (a), which further justifies the validity of VRH scenario since $T_0$ is larger than the measurement temperatures (1.5 K $\leq T \leq$ 45.34 K) of our interest. However, at $T$ below 4.5 K, clear deviation from E-S VRH is observed. As suggested in previous transport studies of inhomogeneous functionalized graphene such as reduced graphene oxide[35] and hydrogenated graphene[36] showing the formation of multi-quantum dots, a substantial reduction of $G$ at low $T$ shown in Fig. 1 (a) can be ascribed to the stochastic Coulomb blockade effect[37, 38] which causes a suppression of current flow. An optical microscopy image of the measured sample in contrast is given in the inset of Fig. 1 (a). Figure 1 (b) shows good fits of the experimental $G(V) = I/V_{sd}$ to Eq. (2) in the specific range of applied source-drain voltage $V_{sd}$ or electric field $E$ at various $T$ ranging from 1.5 K to 3.5 K even though the differential conductance $G(V_{sd} \to 0, T)$ does not conform to the VRH law as a result of blockade effect. This result demonstrates that the applied electric field provides enough energy to carriers to overcome the Coulomb charging energy and the carriers can thereby hop between localized states. The corresponding results for 4.5 K $< T <$ 10 K and for 10 K $< T <$ 20 K are shown in Fig. 2 (a) and Fig. 2 (b), respectively. The $I(V_{sd})$ characteristics at various $T$ are shown in the insets of Fig. 1 (b) and Fig. 2 (a). It can be found that the current is almost suppressed in the bias range of -17 mV $\leq V_{sd} \leq$ 17 mV at $T$ = 1.5 K. We can observe clearly from Fig. 1 (b) that $G$ becomes $T$-independent at high $V_{sd}$, characteristic of activationless hopping transport. In this regime, electric-field-assisted rather than phonon-assisted hopping dominates the transport. In the moderate bias regime where $G(V_{sd})$ is well described by Eq. (2), the energy involved in the hopping process comprises a contribution from phonon absorption and that from electric field. In the low bias regime, deviation from Eq. (2) occurs and becomes significant with increasing $T$ as illustrated in Fig. 2 (a) and Fig. 2 (b).This result is ascribed to the fact that the Ohmic behavior is gradually restored either with increasing $T$ or decreasing $V_{sd}$. In this Ohmic regime, phonon-assisted hopping dominates the transport. Therefore we observe a crossover from phonon-assisted to electric-field assisted VRH with increasing $V_{sd}$. As shown in Fig. 2 (c), the voltage $V_c$, at which the high-field deviation from Eq. (1) begins to occur, shows a $T^{0.85 \pm 0.06}$ dependence, which is in agreement with the criterion $E_{c2} \equiv kT/e\xi$ above which activationless hopping is expected to set in. Activationless behavior does not reveal themselves for 10 K $< T <$ 20 K due to enhanced $E_{c2}$ with increasing $T$ as shown in Fig. 2 (b). From the slope of $\ln G$-versus-$V_{sd}$ dependence at various $T$, we can estimate the length parameter $L$ of Eq. (1) as a function of $T$. Figure 2 (d) shows that $L \propto T^{-x}$ with the exponent $x = 1.24$, which is slightly larger than that predicted by existing theoretical models considering Coulomb interactions. According to Shklovskii model[31], it is expected that $L \propto r_{hop}^2 \propto T^{-2p} = T^{-1}$ with the exponent $p = 1/2$ inferred from $G(V_{sd} \to 0, T)$ in the Ohmic regime at $B = 0$ shown in Fig. 1 (a). However, in the model of Hill[24] and Pollak and Riess[25], $L \propto r_{hop} \propto T^{-p} = T^{-1/2}$ is predicted.

To obtain a thorough understanding of the transport features in our disordered graphene, we perform magneto-transport measurements at $T = 1.5$ K. Figure 3 shows $I$-$V_{sd}$ characteristics at various $B$. The increasing magnitude of $I$ with increasing $B$ suggests that delocalization process occurs as $B$ is applied. However, it occurs only at high enough $V_{sd}$. At low voltage bias, the conducting current is almost vanishing even in the presence of a magnetic field, which can be ascribed to stochastic Coulomb blockade effect. To clarify whether VRH scenario described by Eq. (3) is applicable in the activationless regime, $\ln W$ as a function of $\ln V_{sd}$ at $B = 0$ and 8 T is plotted in the insets (a) and (b) of this figure, respectively, where $W \equiv \partial(\ln I)/\partial V_{sd}$. We note that $\ln W$ depends linearly on $\ln V_{sd}$, indicating that electric-field-assisted VRH is indeed the dominant mechanism for transport in this regime since under this situation $W = \partial(\ln I)/\partial V_{sd} = sV_0^s V_{sd}^{-s-1}$ with $I$ given by Eq. (3) and characteristic $V_0$ related to $E_0$ through a channel length $l = 10.2$ μm. From the slope of the linear fits in the insets of Fig. 3, the exponent $s$ can be determined with high accuracy. At $B = 0$, $s = 0.51 \pm 0.02$ and similarly $s = 0.49 \pm 0.01$ for $B = 8$ T, which further confirms the importance of Coulomb interactions in our disordered graphene in addition to the suggestion given by $T$-dependent data at $T \geq 4.5$ K shown in figure 1 (a). Therefore we plot $\ln I$ against $V_{sd}^{-1/2}$ in Fig. 4 and see that $\ln I$ increases linearly with decreasing $V_{sd}^{-1/2}$ in the specific voltage range. From the slope of this linear relationship between $\ln I$ and $V_{sd}^{-1/2}$, $E_0$ in Eq. (3) is obtained and the localization length $\xi$ can thereby be estimated using the relation $E_0 = V_0/l \sim kT_0/e\xi = 6.2e/4\pi\varepsilon_r\varepsilon_0\xi^2$ with $\varepsilon_r = 2.4$ for graphene[39, 40]. Such results are presented in the inset of Fig. 4 showing that $\xi$ increases with increasing $B$ and demonstrate that the transport carriers are indeed delocalized in the presence of $B$. In strong localization regime, a perpendicular magnetic field can suppress destructive interference among forward scattering paths connecting two hopping sites and thereby results in delocalization of carriers[41-43].

As shown in Fig. 4, we note that at high $V_{sd}$ above 0.18 V in the absence of $B$ the conducting current becomes higher than that predicted by Eq. (3) with $s = 1/2$ for ES VRH, suggesting that delocalization process occurs. Since an electric field can provide energy to carriers, electric-field-assisted delocalization is due to a semi-classical effect different from magnetic-field-driven one discussed above which is fully quantum in nature. With increasing $B$ to 8 T, deviation from Eq. (3) starts at a lower value of $V_{sd} \sim 0.1$ V, which can be associated with the fact that $\xi$ increases with increasing $B$ shown in the inset of Fig. 4. With increasing $\xi$, the carriers acquire more energy from the electric field during each hop and hence deviation can occur at lower $V_{sd}$. One of possible mechanisms that can explain such deviation resorts to Poole-Frenkel effect[44, 45], which argues that the hopping behavior under high electric field condition is strongly relevant to whether donor-like or acceptor-like states such as those induced by hydrogen impurities in our experiment exists and gives an exponential $E$ dependence of current different from that described by Eq. (3). In addition, we cannot rule out the bandgap opening in our hydrogenated graphene as a possible explanation for the observed discrepancy between the experiment and theory at high voltage bias regime. Previous density functional theory calculations[46-48] show that hydrogenated graphene can possess a direct bandgap of ~3.5 eV and semihydrogenatedgraphene have an indirect one of ~0.46 eV. Therefore VRH conduction can proceed through disorder-induced mid-gap states[49-51]. At high $V_{sd}$, the carriers can acquire enough energy to be excited into the propagating band (conduction or valence band)[44], resulting in the breakdown of VRH model as observed in our experiment. To gain more insight into the conduction mechanism beyond activationless-hopping regime, further transport studies of doped graphene are required. We believe that our work presented here provides a comprehensive understanding of transport in highly disordered graphene, which is fundamentally essential to fulfill its potential in graphene-based electronic and optoelectronic devices.

**Conclusion**

In conclusion, we show that a tendency towards activationless hopping transport occurs with increasing source-drain voltage bias $V_{sd}$. At low $V_{sd}$, VRH transport cannot proceed due to blockade effect. At high $V_{sd}$, electric-field-assisted ES VRH dominates the transport, suggesting that Coulomb interactions between charged carriers play an important role in disordered graphene. Moreover we demonstrate that delocalization process occurs in the activationless-hopping regime as a sufficient magnetic field is applied.

**Acknowledgments**

This work was funded by the NSC, Taiwan (grant no: NSC 99-2911-I-002-126), and in part, by National Taiwan University (grant no: 101R7552-2 and 101R890932).

# References

[1] K. S. Novoselov, A. K. Geim, S. V. Morozov, D. Jiang, Y. Zhang, S. V. Dubonos, I. V. Grigorieva, and A. A. Firsov, Science **306**, 666 (2004).

[2] K. S. Novoselov, A. K. Geim, S. V. Morozov, D. Jiang, M. I. Katsnelson, I. V. Grigorieva, S. V. Dubonos, and A. A. Firsov, Nature **438**, 197 (2005).

[3] Y. Zhang, Y.-W. Tan, H. L. Stormer, and P. Kim, Nature **438**, 201 (2005).

[4] C. Lee, X. Wei, J. W. Kysar, and J. Hone, Science **321**, 385 (2008).

[5] A. A. Balandin, S. Ghosh, W. Bao, I. Calizo, D. Teweldebrhan, F. Miao, and C. N. Lau, Nano Lett. **8**, 902 (2008).

[6] K. I. Bolotin, F. Ghahari, M. D. Shulman, H. L. Stormer, and P. Kim, Nature **462**, 196 (2009).

[7] X. Du, I. Skachko, F. Duerr, A. Luican, and E. Y. Andrei, Nature **462**, 192 (2009).

[8] N. Tombros, A. Veligura, J. Junesch, M. H. D. Guimarães, I. J. Vera-Marun, H. T. Jonkman, and B. J. van Wees, Nat. Phys. **7**, 697 (2011).

[9] S. Ihnatsenka and G. Kirczenow, Phys. Rev. B **85**, 121407 (2012).

[10] D. Usachov, O. Vilkov, A. Grüneis, D. Haberer, A. Fedorov, V. K. Adamchuk, A. B. Preobrajenski, P. Dudin, A. Barinov, M. Oehzelt, C. Laubschat, and D. V. Yalikh, Nano Lett. **11**, 5401 (2011).

[11] G. Eda and M. Chhowalla, Adv. Mater. **22**, 2392 (2010).

[12] M. Y. Han, B. Özyilmaz, Y. Zhang, and P. Kim, Phys. Rev. Lett. **98**, 206805 (2007).

[13] P. Gallagher, K. Todd, and D. Goldhaber-Gordon, Phys. Rev. B **81**, 115409 (2010).

[14] B. Standley, A. Mendez, E. Schmidgall, and M. Bockrath, Nano Lett. **12**, 1165 (2012).

[15] N. M. R. Peres, Rev. Mod. Phys. **82**, 2673 (2010).

[16] E. R. Mucciolo and C. H. Lewenkopf, J. Phys.: Condens. Matter **22**, 273201 (2010).

[17] A. H. Castro Neto, F. Guinea, N. M. R. Peres, K. S. Novoselov, and A. K. Geim, Rev. Mod. Phys. **81**, 109 (2009).

[18] H. W. Jiang, C. E. Johnson, and K. L. Wang, Phys. Rev. B **46**, 12830 (1992).

[19] S. I. Khondaker, I. S. Shlimak, J. T. Nicholls, M. Pepper, and D. A. Ritchie, Phys. Rev. B **59**, 4580 (1999).

[20] S.-T. Lo, K. Y. Chen, Y.-C. Su, C. T. Liang, Y. H. Chang, G.-H. Kim, J. Y. Wu, and S.-D. Lin, Solid State Commun. **150**, 1104 (2010).

[21] N. F. Mott, J. Non-Cryst. Solids **1**, 1 (1968).

22. A. L. Efros and B. I. Shklovskii, J. Phys. C **8**, L49 (1975).
23. S. I. Khondaker, I. S. Shlimak, J. T. Nicholls, M. Pepper, and D. A. Ritchie, Solid State Commun. **109**, 751 (1999).
24. R. M. Hill, Philos. Mag. **24**, 1307 (1971).
25. M. Pollak and I. Riess, J. Phys. C **9**, 2339 (1976).
26. N. Apsley and H. P. Hughes, Philos. Mag. **30**, 963 (1974).
27. B. I. Shklovskii, Soviet Physics-Semiconductors **6**, 1964 (1973).
28. D. G. Polyakov and B. I .Shklovskii, Phys. Rev. B **48**, 11167 (1993).
29. S. M. Grannan, A. E. Lange, E. E. Haller, and J. W. Beeman, Phys. Rev. B **45**, 4516 (1992).
30. A. B. Kaiser, C. Gómez-Navarro, R. S. Sundaram, M. Burghard, and K. Kern, Nano Lett. **9**, 1787 (2009).
31. B. I. Shklovskii, Fiz. Tekh. Poluprovodn. **10**, 1440 (1976); [Sov. Phys. Semicond. **10**, 855 (1976)].
32. F. Tremblay, M. Pepper, R. Newbury, D. Ritchie, D. C. Peacock, J. E. F. Frost, G. A. C. Jones, and G. Hill, Phys. Rev. B **40**, 3387 (1989).
33. L. V. Govor, I. A. Bashmakov, K. Boehme, and J. Parisi, J. Appl. Phys. **91**, 739 (2002).
34. C. Chuang, R. K. Puddy, H.-D. Lin, S.-T. Lo, T.-M. Chen, C. G. Smith, and C.-T. Liang, Solid State Commun. **152**, 905 (2012).
35. D. Joung, L. Zhai, and S. I. Khondaker, Phys. Rev. B **83**, 115323 (2011).
36. C. Chuang, R. K. Puddy, M. R. Connoly, S.-T. Lo, H.-D. Lin, T.-M. Chen, C. G. Smith, and C.-T. Liang, Nanoscale Res. Lett. **7**, 459 (2012).
37. I. M. Ruzin, V. Chandrasekhar, E. I. Levin, and L. I. Glazman, Phys. Rev. B **45**, 13469 (1992).
38. M. Kemerink and L. W. Molenkamp, Appl. Phys. Lett. **65**, 1012 (1994).
39. K. W. K. Shung, Phys. Rev. B **34**, 979 (1986).
40. M. C. Lemme, T. J. Echtermeyer, M. Baus, and H. Kurz, IEEE Electron Device Lett. **28**, 282 (2007).
41. F. Tremblay, M. Pepper, D. Ritchie, D. C. Peacock, J. E. F. Frost, and G. A. C. Jones, Phys. Rev. B **39**, 8059 (1989).
42. F. Tremblay, M. Pepper, R. Newbury, D. A. Ritchie, D. C. Peacock, J. E. F. Frost, G. A. C. Jones, and G. Hill, Phys. Rev. B **41**, 8572 (1990).
43. M. E. Raikh and G. F. Wessels, Phys. Rev. B **47**, 15609 (1993).
44. A. K. Jonscher, Thin Solid Films **1**, 213 (1967).
45. A. K. Jonscher, J. Phys. C **4**, 1331 (1971).
46. J. O. Sofo, A. S. Chaudhari, and G. D. Barber, Phys. Rev. B **75**, 153401 (2007).

47   J. Zhou, Q. Wang, Q. Sun, X. S. Chen, Y. Kawazoe, and P. Jena, Nano Lett. **9**, 3867 (2009).
48   J. Zhou, M. M. Wu, X. Zhou, and Q. Sun, Appl. Phys. Lett. **95**, 103108 (2009).
49   D. C. Elias, R. R. Nair, T. M. G. Mohiuddin, S. V. Morozov, P. Blake, M. P. Halsall, A. C. Ferrari, D. W. Boukhvalov, M. I. Katsnelson, A. K. Geim, K. S. Novoselov, Science **323**, 610 (2009).
50   S. -H. Cheng, K. Zou, F. Okino, H. R. Gutierrez, A. Gupta, N. Shen, P. C. Eklund, J. O. Sofo, and J. Zhu, Phys. Rev. B **81**, 205435 (2010).
51   M. Y. Han, J. C. Brant, and P. Kim, Phys. Rev. Lett. **104**, 056801 (2010).

Figure Captions

Figure 1 (a) Logarithm of conductance $G = I/V_{sd}$ as a function of $T^{-1/2}$ in the interval 1.5 K $\leq T \leq$ 45.34 K. The solid line corresponds to a linear fit to Eq. (1) at $T \geq$ 4.5 K. The inset shows the optical microscope image of the sample. (b) ln$G$ as a function of voltage $V_{sd}$ at various temperatures $T$ ranging from 1.5 K to 3.5 K. The solid lines denote the linear dependence of $\ln(I/V_{sd})$ on $V_{sd}$ in the intermediate voltage regime. The inset shows the $I(V_{sd})$ measurements at various $T$. To highlight the suppression of current at low $V_{sd}$, the magnitude of $I(V_{sd})$ at $T$ = 1.5 K is also plotted in the inset.

Figure 2 (a) ln$G$ as a function of $V$ at $T$ = 4.5 K, 5.5 K, 6.5 K, and 10 K. For clarity only data for $I > 0$ and $V_{sd} > 0$ are shown. The inset shows the $I(V_{sd})$ measurements at $T$ = 4.5 K and 20 K. (b) ln$G$ as a function of $V_{sd}$ at various $T$. From top to bottom: $T$ = 20 K, 15 K, 12.5 K, and 10 K, respectively. For clarity only data for both $I > 0$ and $V_{sd} > 0$ are shown. (c) ln$V_c$ versus ln$T$ and (d) ln$L$ as a function of ln$T$.

Figure 3 $I(V_{sd})$ measurements at various magnetic fields $B$. The insets (a) and (b) show ln$W$ as a function of ln$V$ for $B$ = 0 and 8 T, respectively. Here $W \equiv \partial(\ln I)/\partial V_{sd} = sV_0^s V_{sd}^{-s-1}$.

Figure 4 (a) Logarithm of the measured current $I$ as a function of $V_{sd}^{-1/2}$ at $B$ = 0, $B$ = 4 T, and $B$ = 8 T for $T$ = 1.5 K. The black solid lines denote the linear fits to Eq. (3) with $s$ = 1/2. The inset shows the localization length $\xi$ as a function of $B$.

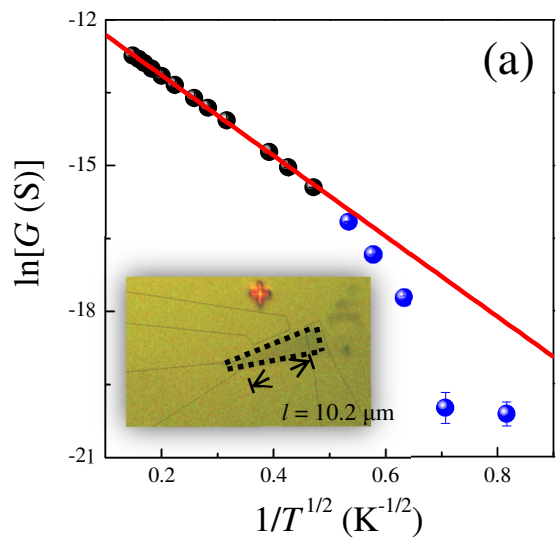

Figure 1 (a) Lo *et al.*

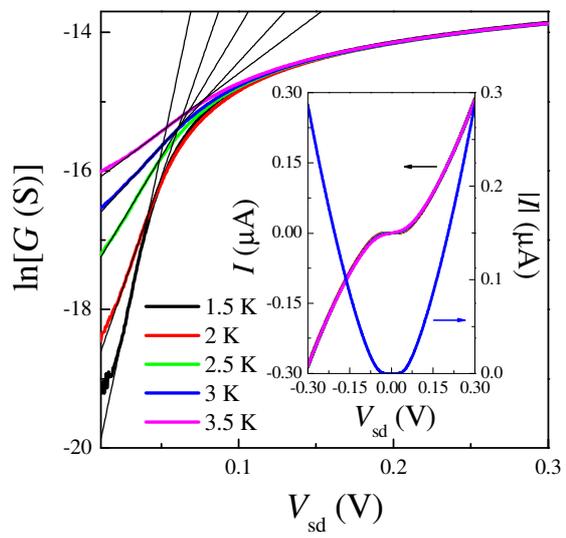

Figure 1 (b) Lo *et al.*

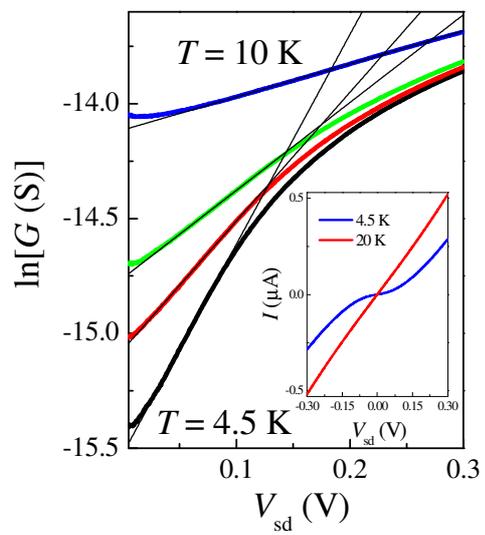

Figure 2 (a) Lo *et al.*

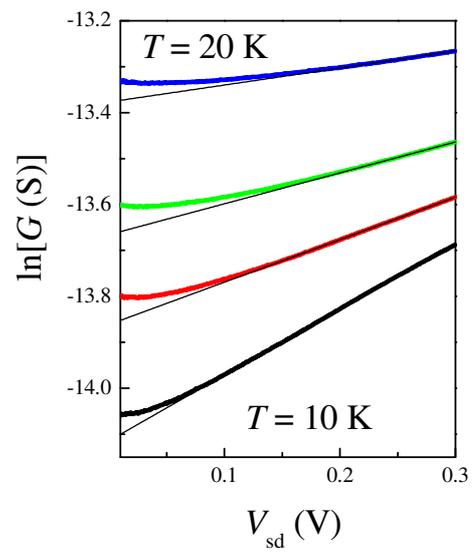

Figure 2 (b) Lo *et al.*

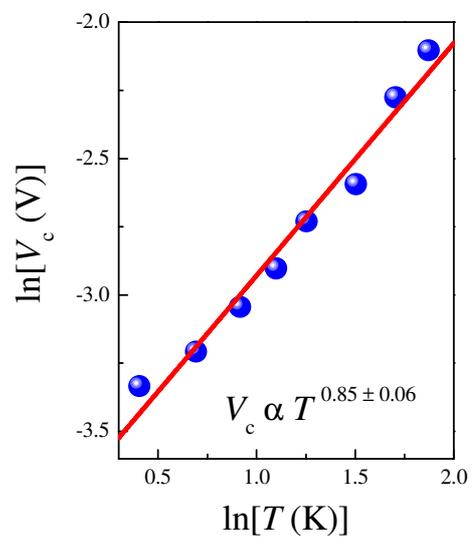

Figure 2 (c) Lo *et al.*

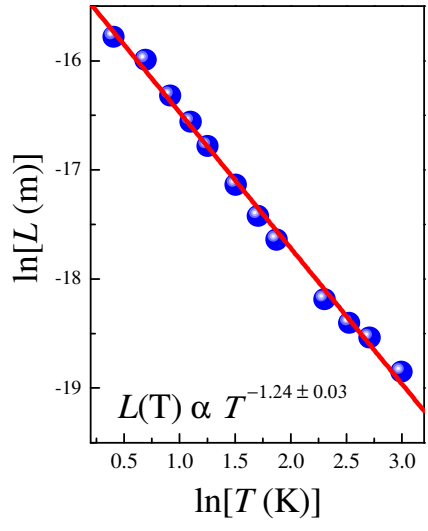

Figure 2 (d) Lo *et al.*

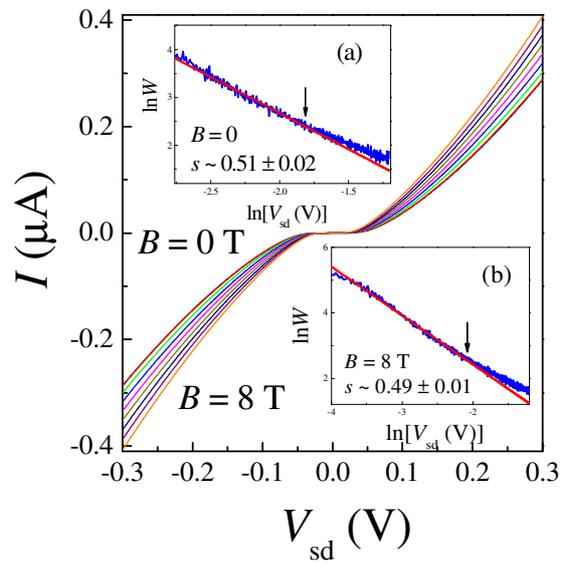

Figure 3 Lo *et al.*

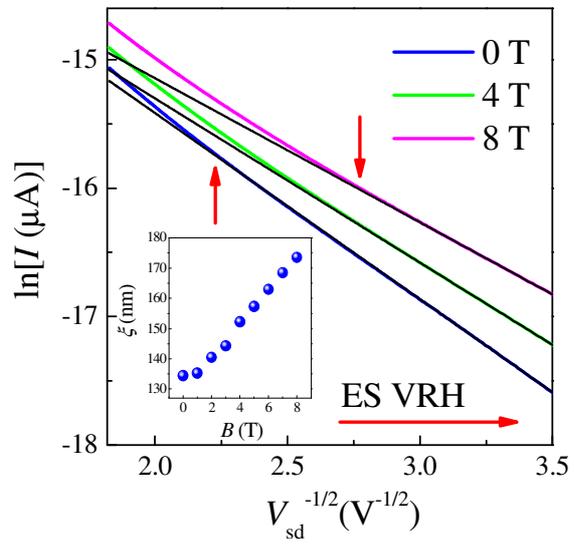

Figure 4 Lo *et al.*